\documentclass[fleqn,usenatbib]{mnras}

\usepackage{newtxtext}
\usepackage[T1]{fontenc}
\DeclareRobustCommand{\VAN}[3]{#2}
\let\VANthebibliography\thebibliography
\def\thebibliography{\DeclareRobustCommand{\VAN}[3]{##3}\VANthebibliography}

\usepackage{graphicx}	
\usepackage{amsmath}	
\usepackage{amssymb}	

\usepackage{newtxmath}
\usepackage{bm,Macro,mathrsfs}

\def\T{\Theta}
\def\tP{\widetilde{P}}

\def\Img{\widetilde{\grad}}
\def\Imx{\widetilde{x}}
\def\hCXX{\widehat{C}^{\rm XX}}

\def\hCYY{\widehat{C}^{\rm YY}}
\def\avecmb#1{\ave{#1}_{\scriptscriptstyle \rm CMB}}

\title[CMB Mode Coupling with Isotropic Polarization Rotation]{CMB Mode Coupling with Isotropic Polarization Rotation}

\author[T. Namikawa]{
Toshiya Namikawa,$^{1}$\thanks{E-mail: tn334@cam.ac.uk}
\\
Department of Applied Mathematics and Theoretical Physics, University of Cambridge, Wilberforce Road, Cambridge CB3 0WA, United Kingdom$^{1}$
}

\pubyear{2021}

\begin{document}
\label{firstpage}
\pagerange{\pageref{firstpage}--\pageref{lastpage}}
\maketitle

\begin{abstract}
We provide a new analysis technique to measure the effect of the isotropic polarization rotation, induced by e.g. the isotropic cosmic birefringence from axion-like particles and a miscalibration of CMB polarization angle, via mode coupling in the cosmic microwave background (CMB). Several secondary effects such as gravitational lensing and CMB optical-depth anisotropies lead to mode coupling in observed CMB anisotropies, i.e., non-zero off-diagonal elements in the observed CMB covariance. To derive the mode coupling, however, we usually assume no parity violation in the observed CMB anisotropies. We first derive a new contribution to the CMB mode coupling arising from parity violation in observed CMB. Since the isotropic polarization rotation leads to parity violation in the observed CMB anisotropies, we then discuss the use of the new mode coupling for constraining the isotropic polarization angle. We find that constraints on the isotropic polarization angle by measuring the new mode-coupling contribution are comparable to that using the $EB$ cross-power spectrum in future high-sensitivity polarization experiments such as CMB-S4 and LiteBIRD. Thus, this technique can be used to cross-check results obtained by the use of the $EB$ cross-power spectrum. 
\end{abstract}


\section{Introduction} \label{sec:intro}

The cosmic microwave background (CMB) anisotropies have been measured precisely over the past decades, and its angular power spectrum has been used to tightly constrain cosmological parameters (e.g. \citealt{P18:main}). Precise measurements of CMB anisotropies also allow us to extract information involved in statistics beyond the angular power spectrum. For example, the gravitational lensing effect on CMB leads to mode coupling between different multipoles of observed CMB anisotropies. Multiple works have developed analysis methods to extract gravitational lensing signals from the mode coupling \citep{HuOkamoto:2001,OkamotoHu:quad}. Using these analysis techniques, multiple CMB observations have successfully measured CMB lensing maps, lensing power spectra (e.g. \citealt{ACT16:phi,BKVIII,P18:phi,PB:phi:2019,SPT:phi:2019}), and cross-power spectra between CMB lensing and other cosmological probes (e.g. \citealt{Larsen:2016,Omori:2018:cmbl-shear,Namikawa:2019:hscxpb,ACT:Omar:2020,ACT:Robertson:2020} and references therein). In addition to lensing, mode coupling in observed CMB anisotropies can be produced by several secondary effects, including the anisotropic polarization rotation, the so-called {\it anisotropic} cosmic birefringence (e.g., \citealt{Kamionkowski:2009:derot,Gluscevic:2012:rot,PB15:rot,BKIX,P16:rot,Namikawa:2020:biref,Bianchini:2021:biref}), the anisotropic CMB optical-depth (e.g., \citealt{Dvorkin:2008:tau,Gluscevic:2012:tau,Namikawa:2017:plktau,Roy:2018:tau,Namikawa:2021:tauxy}), and instrumental and astrophysical systematics \citep{Yadav:2010:inst,Namikawa:2012:bhe,Namikawa:2013:bhepol,Osborne:2013nna,Mishra:2019,Sailer:2020}. 

When deriving the mode coupling, however, we usually assume no parity violation in the observed CMB anisotropies. In the presence of parity violation, the temperature-$B$ and $EB$ cross-power spectra do not vanish. These parity-odd power spectra are further distorted by the above secondary effects, and the observed CMB anisotropies have new mode coupling by parity violation. Extracting this new contribution in mode coupling allows us to extract information on parity violation in the observed CMB anisotropies. 

One of the sources of parity violation is the {\it isotropic} cosmic birefringence in which the CMB linear polarization is rotated by a same angle in all of the line-of-sight directions. Several types of beyond-the-Standard-Model physics can source the isotropic cosmic birefringence such as axion-like particles (ALPs) that couple to photons through a so-called Chern-Simons term (see e.g., \citealt{Carroll:1998,Li:2008,Pospelov:2009,Finelli:2009,Hlozek:2017:axion,Fujita:2020:biref,Fujita:2020:isobiref,Takahashi:2020:biref} and a review, \citealt{Marsh:2016}). The existence of such ALPs is a generic prediction of string theory. In addition, birefringence-inducing pseudo-scalar fields could be candidates for an early dark energy mechanism to resolve the current Hubble parameter tension \citep{Capparelli:2019:CB}. 

Another source of the parity-odd spectra is a miscalibration of the instrumental polarization angle. Calibration of the instrumental polarization angle is accomplished using either hardware calibrators or by measurements of polarized astrophysical sources, but accuracy of these methods is worse than $\mC{O}(0.1)\,$deg \citep{B1rot,PB14:BB,Koopman:2016:optics,P16:rot,BA:2020:rot}. A miscalibration of the polarization angle by $\mC{O}(0.1)\,$deg could produce a non-negligible bias in e.g. measuring the inflationary gravitational waves. To circumvent this situation, in future CMB experiments, \citet{Keating:2013} proposed to use the $EB$ cross-power spectrum to determine the instrumental polarization angle with the assumption that there are no other sources of parity violation (hereafter, the $EB$ spectrum method). 

In this paper, we first derive new CMB mode coupling by the presence of parity violation in the observed CMB anisotropies, i.e., by assuming the non-zero temperature-$B$ and $EB$ parity-odd spectra. We derive a quadratic estimator to extract this new mode coupling analogues to the gravitational lensing. 
If the parity-odd spectra are generated by the polarization rotation, the new mode coupling depends on the polarization angle. 
Thus, we discuss constraining the polarization angle by extracting the new mode coupling via the quadratic estimator and compare its sensitivity to the $EB$ spectrum method. 

This paper is organized as follows. In Sec.~\ref{sec:coupling}, we briefly review the mode coupling induced by the gravitational lensing and then generalize it in the presence of parity violation in the observed CMB anisotropies. In Sec.~\ref{sec:biref}, we consider constraints on the isotropic polarization angle by extracting the mode coupling. Sec.~\ref{sec:discussion} is devoted to discussion and summary. 

\section{CMB mode coupling from parity violation} \label{sec:coupling}

In this section, we briefly summarize the mode coupling arising from lensing and a method to reconstruct lensing map from CMB (for reviews, see, e.g., \citealt{Lewis:2006fu,Hanson:2009:review}). We also discuss other cases of mode coupling. We then derive a new mode coupling contribution in the presence of non-zero temperature-$B$ and $EB$ cross-power spectra. 

\subsection{CMB mode coupling: Gravitational lensing}

The lensed CMB temperature anisotropies, $\tT$, are described as a remapping of the primary CMB anisotropies at the last scattering surface, $\T$, by a deflection vector, $\bm{d}$ \citep{Blanchard:1987AA,Bernardeau:1996aa,Zaldarriaga:1998:LensB}: 
\al{
    \tT(\hatn) 
    = \T(\hatn+\bm{d}(\hatn)) 
    = \T(\hatn) + \bm{d}(\hatn)\cdot\bn\T(\hatn) + \cdots
	\,, 
}
where $\hatn$ is a line-of-sight direction of the observed CMB, $\bn$ is the covariant derivative on the sphere, and in the last equation we expand $\T(\hatn+\bm{d}(\hatn))$ in terms of $\bm{d}(\hatn)$. The deflection vector has two degrees of freedom and is decomposed into the lensing potential and its pseudo-scalar counterpart \citep{Hirata:2003:mle:pol,Cooray:2005:curl}:
\al{
    \bm{d}(\hatn) = \bn\grad(\hatn) + (\star\bn)\curl(\hatn) \,. 
}
In the linear regime with only the scalar perturbations, the lensing potential is given as an integral of the gravitational potential along the photon geodesic while the pseudo-scalar lensing potential, $\curl$, vanishes (e.g. \citealt{Yamauchi:2012:vector,Yamauchi:2013:TAM}) \footnote{The curl mode is usually ignored since its contribution is negligible in the standard scenario where the curl mode is generated by the non-linear scalar density perturbations \citep{Saga:2015}. In this paper, therefore, we ignore the curl mode signals.}
Similarly, denoting the Stokes Q and U parameters as $Q$ and $U$ and defining $P=Q+\iu U$, the lensed CMB polarization anisotropies, $\tP$, are given by: 
\al{
    \tP(\hatn) = P(\hatn+\bm{d}(\hatn)) 
    = P(\hatn) + \bm{d}(\hatn)\cdot\bn P(\hatn) + \cdots
	\,. 
}
We define the spherical harmonic coefficients of a spin-$0$ quantity, $x$, as:
\al{
    x_{\l m} = \Int{2}{\hatn}{} Y^*_{\l m}(\hatn) x(\hatn)
    \,. 
}
We also define the rotationally invariant $E$- and $B$-modes of the CMB polarization from the Stokes parameters \citep{Zaldarriaga:1996:EBdef}:
\al{
    E_{\l m} \pm\iu B_{\l m} = - \Int{2}{\hatn}{} (Y_{\l m}^2(\hatn))^* [Q\pm\iu U](\hatn)
    \,. 
}
We then have the expression for the lensed CMB anisotropies up to linear order in $\grad$ and $\curl$ as:
\al{
	\tT_{\l m} &= \T_{\l m} + \sum_{LM\l'm'}(-1)^m\Wjm{\l}{L}{\l'}{-m}{M}{m'}  
	\notag \\
	&\qquad \times \sum_{x=\grad,\curl}x_{LM}\T_{\l'm'}W^{x,0}_{\l L\l'} 
	\,, \label{Eq:T} \\ 
	\tE_{\l m} &= E_{\l m} + \sum_{LM\l'm'}(-1)^m\Wjm{\l}{L}{\l'}{-m}{M}{m'}
	\notag \\
	&\qquad \times \sum_{x=\grad,\curl}x_{LM}
	\left(E_{\l'm'}W^{x,+}_{\l L\l'} + B_{\l'm'}W^{x,-}_{\l L\l'}\right) 
	\,, \label{Eq:E} \\ 
	\tB_{\l m} &= B_{\l m} + \sum_{LM\l'm'}(-1)^m\Wjm{\l}{L}{\l'}{-m}{M}{m'} 
	\notag \\
	&\qquad \times \sum_{x=\grad,\curl}x_{LM}
	\left(-E_{\l'm'}W^{x,-}_{\l L\l'} + B_{\l'm'}W^{x,+}_{\l L\l'}\right) 
	\,, \label{Eq:B}
}
where we define:
\al{
    W^{x,+}_{\l L\l'} 
	&\equiv \frac{W^{x,+2}_{\l L\l'}+W^{x,-2}_{\l L\l'}}{2} 
	\,, \\ 
    W^{x,-}_{\l L\l'} 
	&\equiv \iu\frac{W^{x,+2}_{\l L\l'}-W^{x,-2}_{\l L\l'}}{2} 
    \,, 
}
with:
\al{
	&(-1)^m\Wjm{\l}{L}{\l'}{-m}{M}{m'} W^{\grad,\pm 2}_{\l L\l'} 
	= \Int{2}{\hatn}{} (Y^{\pm 2}_{\l m})^* \bn Y_{LM} \cdot \bn Y^{\pm 2}_{\l'm'}
	\\ 
	&(-1)^m\Wjm{\l}{L}{\l'}{-m}{M}{m'} W^{\curl,\pm 2}_{\l L\l'} 
	= \Int{2}{\hatn}{} (Y^{\pm 2}_{\l m})^* (\star\bn) Y_{LM} \cdot \bn Y^{\pm 2}_{\l'm'}
	\,. 
}
The above lensed $E$- and $B$-modes introduce the following mode coupling \citep{OkamotoHu:quad,Namikawa:2011:curlrec}: 
\al{
	\avecmb{\tX_{\l m}\tY_{\l'm'}}
	= \sum_{x=\grad,\curl}\sum_{LM}\Wjm{\l}{\l'}{L}{m}{m'}{M} f^{x,({\rm XY})}_{\l L\l'} x^*_{LM} 
	\,, \label{Eq:kernel} 
}
where $\avecmb{\cdots}$ denotes the ensemble average over the primary CMB anisotropies with a fixed realization of the distortion fields. We ignore the higher-order terms of the distortion fields. The kernel functions, $f^{x,(\rm XY)}_{\l L\l'}$, which characterize the mode coupling are given as:
\al{
	f^{x,(\T\T)}_{\l L\l'} &= W^{x,0}_{\l L\l'}\CTT_{\l'} + p_{\l L\l'} W^{x,0}_{\l'L\l}\CTT_\l 
	\,, \label{Eq:kernel:TT} \\ 
	f^{x,(\T E)}_{\l L\l'} &= W^{x,0}_{\l L\l'}\CTE_{\l'} + p_{\l L\l'} W^{x,+}_{\l'L\l}\CTE_\l 
	\,, \label{Eq:kernel:TE} \\ 
	f^{x,(\T B)}_{\l L\l'} &= -p_{\l L\l'}W^{x,-}_{\l'L\l}\CTE_\l 
	\,, \label{Eq:kernel:TB} \\ 
	f^{x,(EE)}_{\l L\l'}   &= W^{x,+}_{\l L\l'}\CEE_{\l'} + p_{\l L\l'} W^{x,+}_{\l'L\l}\CEE_\l 
	\,, \label{Eq:kernel:EE} \\ 
	f^{x,(EB)}_{\l L\l'}   &= W^{x,-}_{\l L\l'}\CBB_{\l'} - p_{\l L\l'} W^{x,-}_{\l'L\l}\CEE_\l 
	\,, \label{Eq:kernel:EB} \\ 
	f^{x,(BB)}_{\l L\l'}   &= W^{x,+}_{\l L\l'}\CBB_{\l'} + p_{\l L\l'} W^{x,+}_{\l'L\l}\CBB_\l 
	\,. \label{Eq:kernel:BB}
}
With a quadratic combination of the observed CMB anisotropies, $\hX$ and $\hY$, the quadratic estimator is given by \citep{OkamotoHu:quad}:
\al{
	(\estx^{\rm XY}_{LM})^* = \frac{A^{x,({\rm XY})}_L}{\Delta^{\rm XY}} 
	\sum_{\l\l'mm'}\Wjm{\l}{\l'}{L}{m}{m'}{M}
	(f^{x,({\rm XY})}_{\l L\l'})^*\frac{\hX_{\l m}}{\hCXX_\l}\frac{\hY_{\l'm'}}{\hCYY_\l}
	\,, \label{Eq:estg} 
}
where $\Delta^{\rm XX}=2$, $\Delta^{\rm EB}=\Delta^{\rm TB}=1$, and $A^{x,({\rm XY})}_L$ is the estimator normalization so that the above estimator is unbiased: 
\al{
	\left[A^{x,({\rm XY})}_L\right]^{-1}
	= \frac{1}{2L+1}\sum_{\l\l'}\frac{\left|f^{x,({\rm XY})}_{\l L\l'}\right|^2}{\Delta^{\rm XY}\hCXX_\l\hCYY_{\l'}} 
	\,, \label{Eq:Norm}
}
with $\hCXX_\l$ ($\hCYY_\l$) denoting the observed power spectrum. The angular power spectrum of the reconstruction noise is given by \eq{Eq:Norm} in the idealistic case \citep{OkamotoHu:quad}.

\subsection{CMB mode coupling: Other sources}

Mode coupling is led by not only the gravitational lensing but also the anisotropies in CMB optical depth, anisotropic cosmic birefringence, observational effects such as survey window, masking and inhomogeneous noise, and extragalactic foregrounds. 
If multiple effects introduce mode coupling, the right-hand-side of \eq{Eq:kernel} becomes the sum of all these effects:
\al{
	\sum_x\sum_{LM}\Wjm{\l}{\l'}{L}{m}{m'}{M} f^{x,({\rm XY})}_{\l L \l'} x^*_{LM} 
	\,, \label{Eq:mode-coupling:multiple} 
}
where $x$ denotes a field causing the mode coupling, e.g., the anisotropic birefringence angle, anisotropic CMB optical depth, lensing potential, $\grad$, and curl mode $\curl$. The functional form of $f^{x,(\rm XY)}_{\l L\l'}$ depends on the field, $x$. 


In general, we can decompose mode coupling terms into two groups in terms of parity symmetry; one has the same parity symmetry as the lensing potential such as masking and point sources (parity even), and the other has the opposite parity symmetry to the lensing potential such as the curl mode and anisotropic cosmic birefringence (parity odd). 
Analogues to the simultaneous estimate of the lensing potential and curl mode \citep{Namikawa:2011:curlrec}, the difference of the parity symmetry allows us to separately estimate the mode coupling terms from the parity even and parity odd groups. 

\subsection{CMB mode coupling from polarization rotation}

Next, we generalize the mode coupling described in the previous section to the case when parity violation exists in the observed CMB anisotropies. As an example, we first consider a case in the presence of the isotropic polarization rotation in which the observed polarization is given by:
\al{
    P'(\hatn) = \E^{2\iu\beta}\tP(\hatn) \,. \label{Eq:P:isorot}
}
Here, $\tP$ is the lensed CMB anisotropies, and $\beta$ is the isotropic polarization angle which does not depend on observed line-of-sight directions. 

For example, $\beta$ equals to a miscalibrated polarization angle. The polarization angle, $\beta$, is also interpreted as the isotropic cosmic birefringence angle induced by the ALPs as follows. We first note that, in the absence of lensing, the isotropic cosmic birefringence effect on CMB by the ALPs is given by: 
\al{
    P^{\rm rot}(\hatn) = \E^{2\iu\beta}P(\hatn) \,, \label{Eq:P:isorot:0}
}
where $P$ denotes the primary CMB polarization and the birefringence angle is given by $\beta=g_{a\gamma}\Delta a/2$ \citep{Carroll:1989:rot,Harari:1992:axion}. Here, $\Delta a$ is the change in the ALP fields over the photon trajectory, and $g_{\alpha\gamma}$ is the coupling constant between the ALPs and photons. We focus on the background evolution of the ALPs which only depends on time so that $\beta$ is isotropic. 
In the presence of lensing, the path of CMB photons is deflected and the CMB polarization is rotated along the deflected path. If the ALP fields do not depend on spatial coordinates, $\beta$ does not depend on the path of CMB photons, i.e., observed CMB anisotropies at each direction are rotated by the same angle, $\beta$, even in the presence of lensing. 
The two operations, the rotation and lensing, commute, 
\footnote{
\eq{Eq:P:isorot} means that we first remap the primary polarization map and then rotate the lensed polarization map. On the other hand, one may consider a case where we first rotate the primary polarization map and then remap the rotated polarization map, $P^{\rm rot}(\hatn)$. In the isotropic birefringence case, the remapping only operates on to the $P(\hatn)$ in \eq{Eq:P:isorot:0}, yielding \eq{Eq:P:isorot}. 
} 
and the observed CMB polarization is given by \eq{Eq:P:isorot}. 

Eq.~\eqref{Eq:P:isorot} yields (e.g. \citealt{Zhao:2014:biref}):
\al{
    E'_{\l m} &= \tilde{E}_{\l m}\cos2\beta - \tilde{B}_{\l m}\sin2\beta
    \,, \label{Eq:E:new} \\ 
    B'_{\l m} &= \tilde{B}_{\l m}\cos2\beta + \tilde{E}_{\l m}\sin2\beta
    \,. \label{Eq:B:new}
}
We assume $\beta\ll 1$. Expanding \eq{Eq:E:new,Eq:B:new} to linear order in $\beta$, we obtain:
\al{
    E'_{\l m} &\simeq \tilde{E}_{\l m} - 2\beta\tilde{B}_{\l m}
    \,, \label{Eq:E:new:expand} \\ 
    B'_{\l m} &\simeq \tilde{B}_{\l m} + 2\beta\tilde{E}_{\l m}
    \,. \label{Eq:B:new:expand}
}
Substituting the above equations into \eq{Eq:mode-coupling:multiple}, the off-diagonal covariance of the distorted CMB anisotropies has the following new contributions induced by the isotropic polarization rotation:
\al{
    &\avecmb{E'_{\l m}E'_{\l'm'}} 
    \supset - 2\beta[\avecmb{\tE_{\l m}\tB_{\l'm'}} + \avecmb{\tB_{\l m}\tE_{\l'm'}}]
    \notag \\
    &\quad= \sum_{LM}\Wjm{\l}{\l'}{L}{m}{m'}{M} 
    \left[-2(f^{\grad,\rm EB}_{\l L\l'}+p_{\l L\l'}f^{\grad,\rm EB}_{\l'L\l})\Img^*_{LM} \right]  
    \,, \label{Eq:EE:lens-beta} \\
    &\avecmb{E'_{\l m}B'_{\l'm'}} 
    \supset 2\beta[\avecmb{\tE_{\l m}\tE_{\l'm'}}-\avecmb{\tB_{\l m}\tB_{\l'm'}}]
    \notag \\
    &\quad= \sum_{LM}\Wjm{\l}{\l'}{L}{m}{m'}{M} 
    \left[2(f^{\grad,\rm EE}_{\l L\l'}-f^{\grad,\rm BB}_{\l L\l'})\Img^*_{LM}\right] 
    \,, \label{Eq:EB:lens-beta} \\
    &\avecmb{B'_{\l m}B'_{\l'm'}} 
    \supset 2\beta[\avecmb{\tB_{\l m}\tE_{\l'm'}}+\avecmb{\tE_{\l m}\tB_{\l'm'}}]
    \notag \\
    &\quad= \sum_{LM}\Wjm{\l}{\l'}{L}{m}{m'}{M} 
    \left[2(f^{\grad,\rm EB}_{\l L\l'}+p_{\l L\l'}f^{\grad,\rm EB}_{\l'L\l}) \Img^*_{LM} \right] 
    \,, \label{Eq:BB:lens-beta}
}
where we define $\Img=\beta \grad$. We ignore the higher order terms, $\mC{O}(\beta^2)$. Note that the temperature anisotropies are not affected by the rotation, i.e., $\T'=\tT$. From \eq{Eq:E:new:expand,Eq:B:new:expand,Eq:mode-coupling:multiple}, we find
\al{
    \avecmb{\T'_{\l m}E'_{\l'm'}} 
    &\supset -2\beta\avecmb{\tT_{\l m}\tB_{\l'm'}}
    \notag \\ 
    &= \sum_{LM}\Wjm{\l}{\l'}{L}{m}{m'}{M} 
    \left(-2f^{\grad,\rm \T B}_{\l L\l'} \Img^*_{LM} \right) 
    \,, \label{Eq:TE:lens-beta} \\ 
    \avecmb{\T'_{\l m}B'_{\l'm'}} 
    &\supset 2\beta\avecmb{\tT_{\l m}\tE_{\l'm'}}
    \notag \\
    &= \sum_{LM}\Wjm{\l}{\l'}{L}{m}{m'}{M} 
    \left( 2f^{\grad,\rm \T E}_{\l L\l'}\Img^*_{LM} \right)
    \,, \label{Eq:TB:lens-beta} 
}
Using \eq{Eq:mode-coupling:multiple} and the explicit expressions for the mode-coupling kernel function of the lensing distortion (\eq{Eq:kernel:TT,Eq:kernel:TE,Eq:kernel:TB,Eq:kernel:EE,Eq:kernel:EB,Eq:kernel:BB}), we obtain the kernel functions for the mode coupling induced by $\Img$:
\al{
    f^{\Img,(\rm \T E)}_{\l L\l'} &= -2f^{\grad,(\rm \T B)}_{\l L\l'} 
    = 2p_{\l L\l'}W^{\grad,-}_{\l'L\l}\CTE_\l
    \,, \label{Eq:kernel:TE:Img} \\
    f^{\Img,(\rm \T B)}_{\l L\l'} &= 2f^{\grad,(\rm \T E)}_{\l L\l'} 
    = 2W^{\grad,0}_{\l L\l'}\CTE_{\l'} + 2p_{\l L\l'} W^{\grad,+}_{\l'L\l}\CTE_\l 
    \,, \label{Eq:kernel:TB:Img} \\
    f^{\Img,(\rm EE)}_{\l L\l'} &= -2(f^{\grad,(\rm EB)}_{\l L\l'}+p_{\l L\l'}f^{\grad,(\rm EB)}_{\l'L\l}) 
    \notag \\ 
    &= 2[W^{\grad,-}_{\l L\l'}(\CEE_{\l'}-\CBB_{\l'})+p_{\l L\l'}W^{\grad,-}_{\l'L\l}(\CEE_\l-\CBB_\l)]
    \,, \label{Eq:kernel:EE:Img} \\
    f^{\Img,(\rm EB)}_{\l L\l'} &= 2(f^{\grad,(\rm EE)}_{\l L\l'}-f^{\grad,(\rm BB)}_{\l L\l'}) 
    \notag \\ 
    &= 2W^{\grad,+}_{\l L\l'}(\CEE_{\l'}-\CBB_{\l'}) + 2p_{\l L\l'} W^{\grad,+}_{\l'L\l}(\CEE_\l-\CBB_\l)
    \,, \label{Eq:kernel:EB:Img} \\
    f^{\Img,(\rm BB)}_{\l L\l'} &= 2(f^{\grad,(\rm EB)}_{\l L\l'}+p_{\l L\l'}f^{\grad,(\rm EB)}_{\l'L\l})
    = - f^{\Img,(\rm EE)}_{\l L\l'}
    \,. \label{Eq:kernel:BB:Img} 
}
Note that $f^{\Img,(\rm XY)}_{\l L\l'}$ is orthogonal to $f^{\grad,(\rm XY)}_{\l L\l'}$, i.e.:
\al{
    f^{\grad,(\rm XY)}_{\l L\l'}f^{\Img,(\rm XY)}_{\l L\l'} = 0 
    \,, \label{Eq:f-parity}
}
for any combination of $\l,L,\l'$. This property comes from the fact that the distortion fields, $\grad$ and $\Img$, have even and odd parity symmetry, respectively. This relationship is similar to that between the lensing potential and curl mode, and we can estimate $\grad_{LM}$ and $\Img_{LM}$ separately using the distinctive property in parity symmetry \citep{Namikawa:2011:curlrec}.

\subsection{Parity violation in observed CMB anisotropies}

Here, we generalize the kernel functions of \eq{Eq:kernel:TE:Img,Eq:kernel:TB:Img,Eq:kernel:EE:Img,Eq:kernel:EB:Img,Eq:kernel:BB:Img} to the case in the presence of the non-zero $\T B$ and $EB$ spectra. 

The kernel functions with the odd-parity spectra are derived by substituting \eq{Eq:T,Eq:E,Eq:B} into \eq{Eq:kernel} but keeping the $\T B$ and $EB$ spectra. Denoting $x$ as a source of mode coupling (e.g., $x=\phi$), we find that the kernel functions arising from the non-zero odd-parity spectra are given by: 
\al{
    f^{\Imx,(\rm \T E)}_{\l L\l'} 
    &= p_{\l L\l'} W^{x,-}_{\l'L\l}\CTB_\l
    \,, \label{Eq:kernel:TE:Imx} \\
    f^{\Imx,(\rm \T B)}_{\l L\l'} 
    &= W^{x,0}_{\l L\l'}\CTB_{\l'} + p_{\l L\l'} W^{x,+}_{\l'L\l}\CTB_\l
    \,, \label{Eq:kernel:TB:Imx} \\
    f^{\Imx,(\rm EE)}_{\l L\l'} 
    &= W^{x,-}_{\l L\l'}\CEB_{\l'} + p_{\l L\l'} W^{x,-}_{\l'L\l}\CEB_\l 
    \,, \label{Eq:kernel:EE:Imx} \\
    f^{\Imx,(\rm EB)}_{\l L\l'} 
    &= W^{x,+}_{\l L\l'}\CEB_{\l'} + p_{\l L\l'} W^{x,+}_{\l'L\l}\CEB_\l
    \,, \label{Eq:kernel:EB:Imx} \\
    f^{\Imx,(\rm BB)}_{\l L\l'} 
    &= -W^{x,-}_{\l L\l'}\CEB_{\l'} - p_{\l L\l'}W^{x,-}_{\l'L\l}\CEB_\l 
    \,, \label{Eq:kernel:BB:Imx}
}
where \eq{Eq:kernel} has the following new term:
\al{
	\avecmb{X'_{\l m}Y'_{\l'm'}}
	&= \sum_x\sum_{LM}\Wjm{\l}{\l'}{L}{m}{m'}{M} 
	\notag \\
	&\qquad\times [f^{x,({\rm XY})}_{\l L \l'}+f^{\Imx,({\rm XY})}_{\l L \l'}] x^*_{LM} 
	\,. \label{Eq:mode-coupling:multiple:general} 
}
Note that, substituting $\CTB=2\beta\CTE$ and $\CEB=2\beta(\CEE-\CBB)$ into the above kernel functions, dividing the kernel functions by $\beta$, and choosing $x=\grad$, we reproduce the same kernel function as that obtained in \eq{Eq:kernel:TE:Img,Eq:kernel:TB:Img,Eq:kernel:EE:Img,Eq:kernel:EB:Img,Eq:kernel:BB:Img}. 

Finally, the quadratic estimator of $\Imx$ is the same as that of $x$ but with the kernel functions given by \eq{Eq:kernel:TE:Imx,Eq:kernel:TB:Imx,Eq:kernel:EE:Imx,Eq:kernel:EB:Imx,Eq:kernel:BB:Imx} in \eq{Eq:estg,Eq:Norm}. Due to the property of parity symmetry, \eq{Eq:f-parity} holds for $x$ and $\Imx$. This means that the quadratic estimator of $\Imx$ is not biased by the presence of $x$ at the leading order of $x$.

\section{Application to constraining isotropic polarization angle} \label{sec:biref}

In this section, we discuss constraining the isotropic polarization angle by reconstructing $\Img=\beta\grad$ with the estimator described in the previous section.

\subsection{Reconstruction with the quadratic estimator}

\begin{figure}
\centering
\includegraphics[width=85mm,clip]{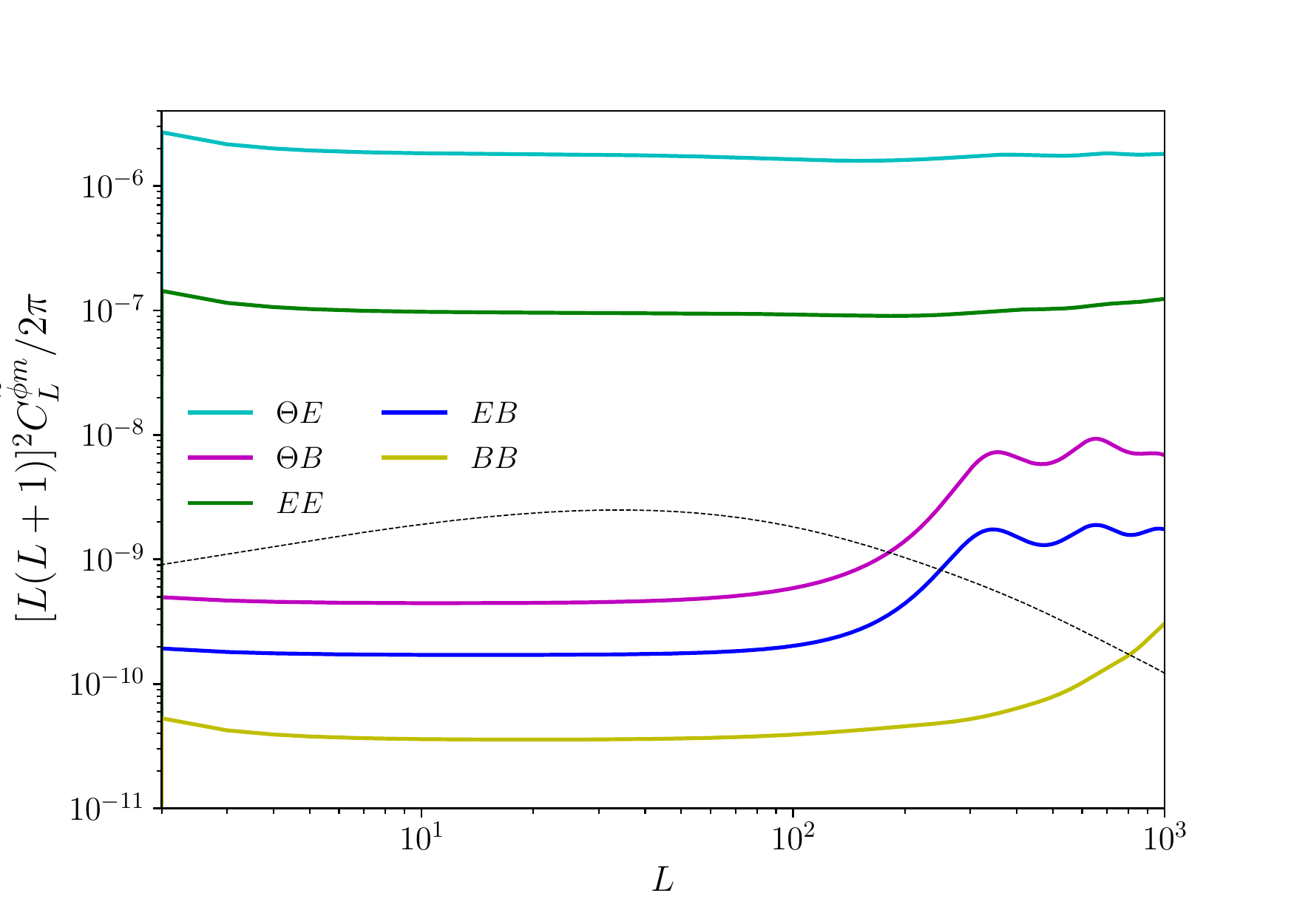}
\caption{
The reconstruction noise power spectra of $\Img$ from each quadratic estimator. We assume the $5\,\mu$K' white noise for polarization with temperature noise $\sqrt{2}$ lower and the $1$ arcmin FWHM Gaussian beam. We use the CMB multipoles up to $4000$ for the reconstruction. The black line shows the cross-power spectrum between $\Img$ and a perfectly correlated mass tracer, i.e., the true CMB lensing potential, with $\beta=1$\,deg. 
}
\label{fig:norm}
\end{figure}

\begin{figure}
\centering
\includegraphics[width=85mm,clip]{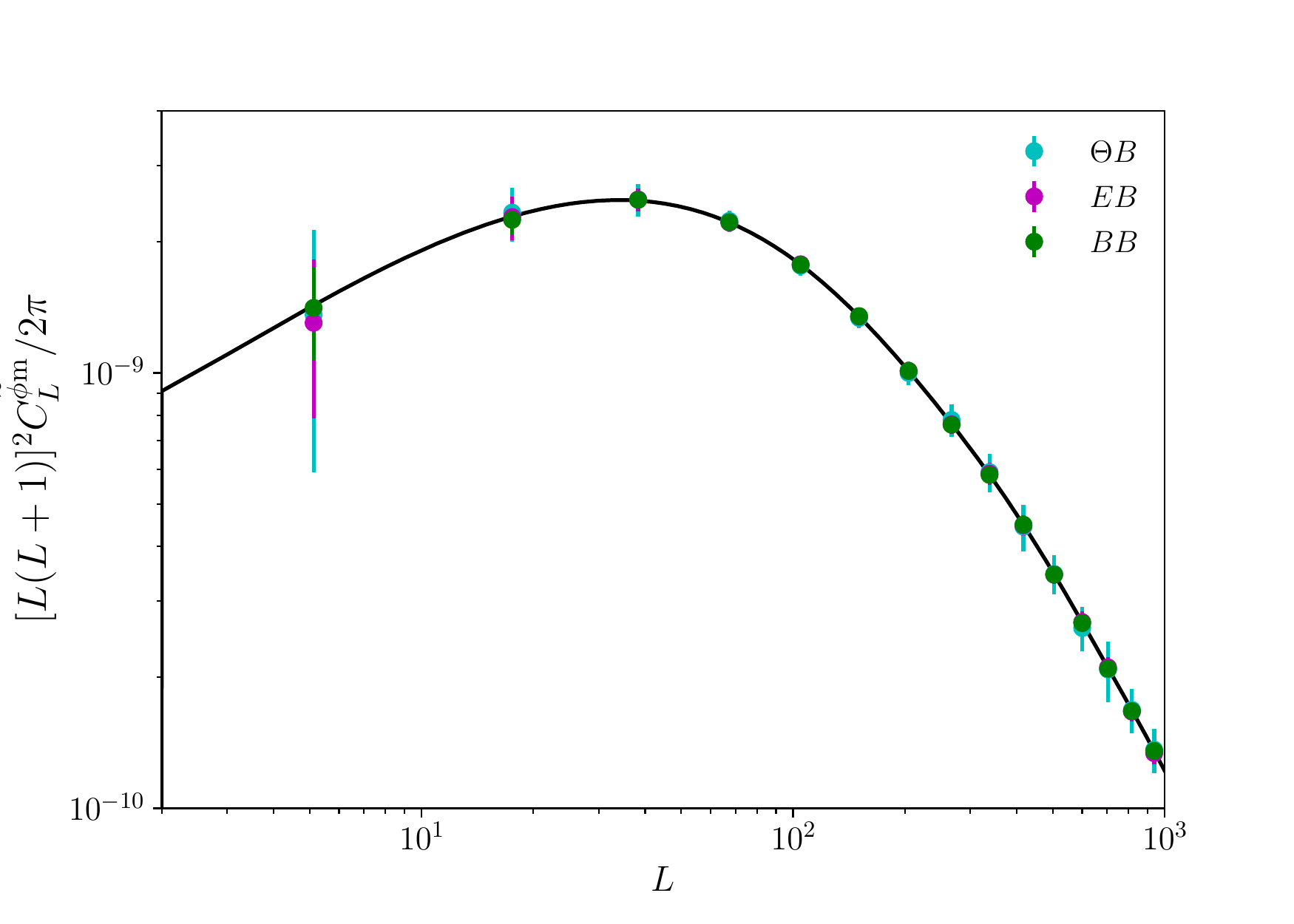}
\caption{
The cross-power spectrum between $\Img$ and a perfectly correlated mass tracer, i.e., the true CMB lensing potential. We reconstruct $\Img$ using the $\T B$, $EB$ or $BB$ estimators. The mean and error bars are obtained from $100$ realizations. We assume $\beta=1$\,deg, the $5\,\mu$K-arcmin white noise level for polarization with temperature noise $\sqrt{2}$ lower and the $1$ arcmin FWHM Gaussian beam for a full sky observation. The black solid line shows the expected signal cross-power spectrum, $\beta C^{\grad\grad}_L$. The good agreement between the simulated and expected cross-power spectrum verifies that we can robustly measure the cross-power spectrum using the estimator of $\Img$. 
}
\label{fig:sim}
\end{figure}

We first check the reconstruction noise of the quadratic estimators given by \eq{Eq:kernel:TE:Img,Eq:kernel:TB:Img,Eq:kernel:EE:Img,Eq:kernel:EB:Img,Eq:kernel:BB:Img}. 
Fig.~\ref{fig:norm} shows the reconstruction noise power spectrum defined in \eq{Eq:Norm} for each quadratic estimator. We assume a white noise level of $5\,\mu$K-arcmin for polarization, with temperature noise $\sqrt{2}$ lower, and the $1$ arcmin FWHM Gaussian beam. Note that, here and after, we replace the primary unlensed CMB spectra to the lensed CMB spectra in the kernel functions analogues to \citet{Hanson:2010:N2} in order to mitigate the higher-order biases. In this experimental setup, the most efficient estimator is the $BB$ estimator. This is because the $BB$ estimator contains the mode coupling induced by the $EB$ spectrum while its reconstruction noise level is determined by the $B$-modes and is not limited by the cosmic variance of $E$-modes. The reconstruction noise spectrum of the $BB$ estimator depends on the forth power of the $B$-modes. Thus, if the polarization noise level becomes lower, the reconstruction noise becomes reduced rapidly unless the observed $B$-modes are dominated by the lensing $B$-modes. 

Next, we use a Monte Carlo simulation to test whether $\Img$ is accurately reconstructed from the quadratic estimator using the weight functions of \eq{Eq:kernel:TE:Img,Eq:kernel:TB:Img,Eq:kernel:EE:Img,Eq:kernel:EB:Img,Eq:kernel:BB:Img}. 
We generate fullsky primary CMB anisotropies and lensing potential maps as random Gaussian fields. Fullsky lensed CMB maps are obtained by remapping the primary CMB anisotropies with the lensing potential. The lensed CMB polarization maps are further rotated by $\beta=1$\,deg to obtain observed CMB polarization maps described in \eq{Eq:P:isorot}. The $5\,\mu$K-arcmin and $5/\sqrt{2}\,\mu$K-arcmin white noise are added to the polarization and temperature maps, respectively, and we assume a $1$ arcmin FWHM Gaussian beam. 
Fig.~\ref{fig:sim} shows an example of measuring cross-power spectrum between a reconstructed $\Img$ and the input lensing potential, $\grad$. If the estimator works correctly, the cross-power spectrum equals to $\beta C^{\grad\grad}_L$. The cross-power spectrum between $\Img$ and $\grad$ obtained from the simulation is in good agreement with $\beta C^{\grad\grad}_L$. Thus, the reconstruction method derived in Sec.~\ref{sec:coupling} provides a method to robustly measure the cross-power spectrum.

\subsection{Sensitivity to isotropic polarization angle}

\begin{figure}
\centering
\includegraphics[width=85mm]{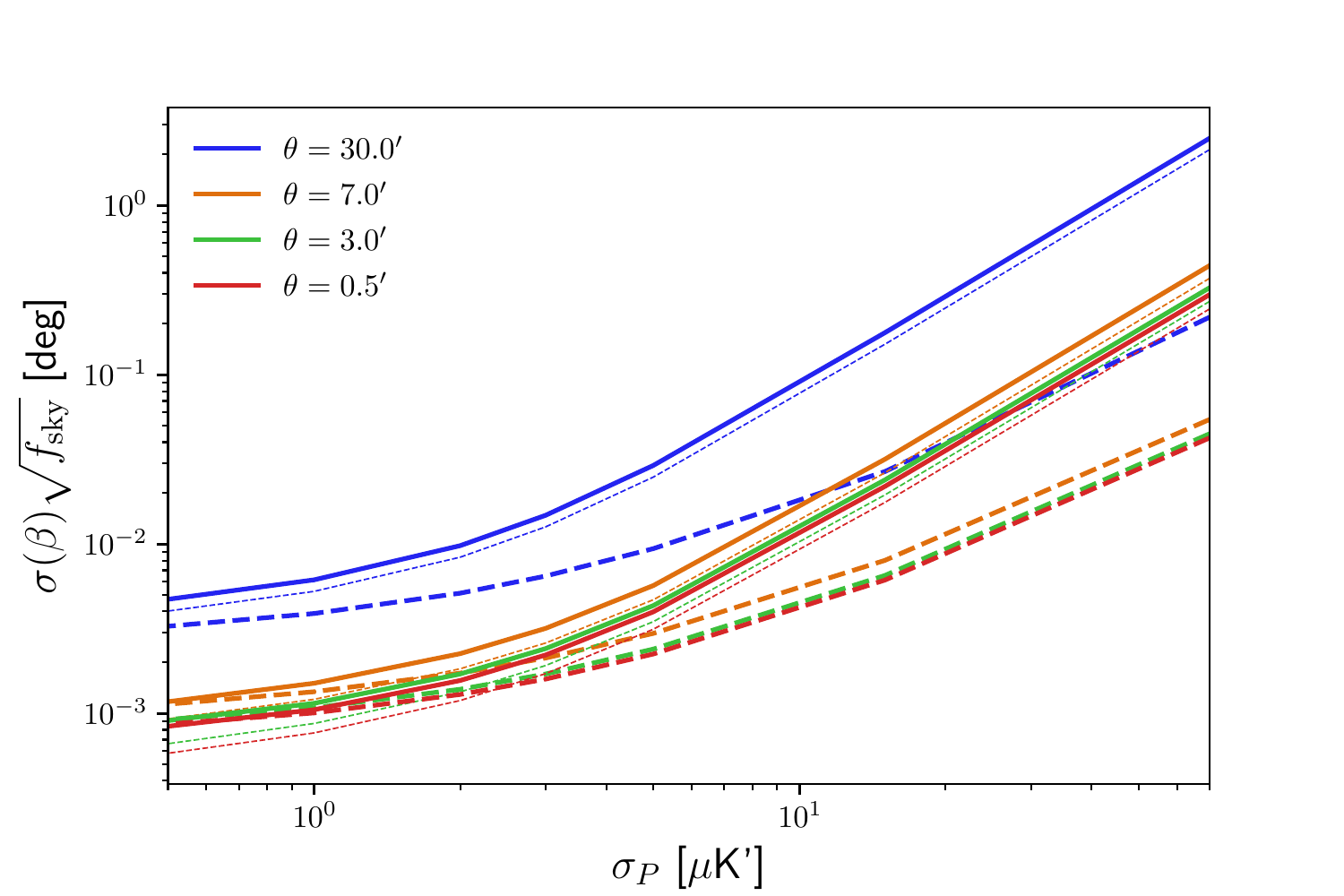}
\caption{
Expected $1\,\sigma$ constraints on the polarization angle as a function of the white noise level in polarization and FWHM beam size. The solid lines show the case using the cross-correlation between $\Img$ and CIB, while the dashed lines show the case with the $EB$ spectrum method. We adopt the correlation coefficient between the CIB and lensing potential provided by \citet{Sherwin:2015}. The thin dotted lines show the case if we use a perfectly correlated mass tracer instead of the CIB. We do not include foregrounds. 
}
\label{fig:comp}
\end{figure}

\begin{figure}
\centering
\includegraphics[width=85mm,clip]{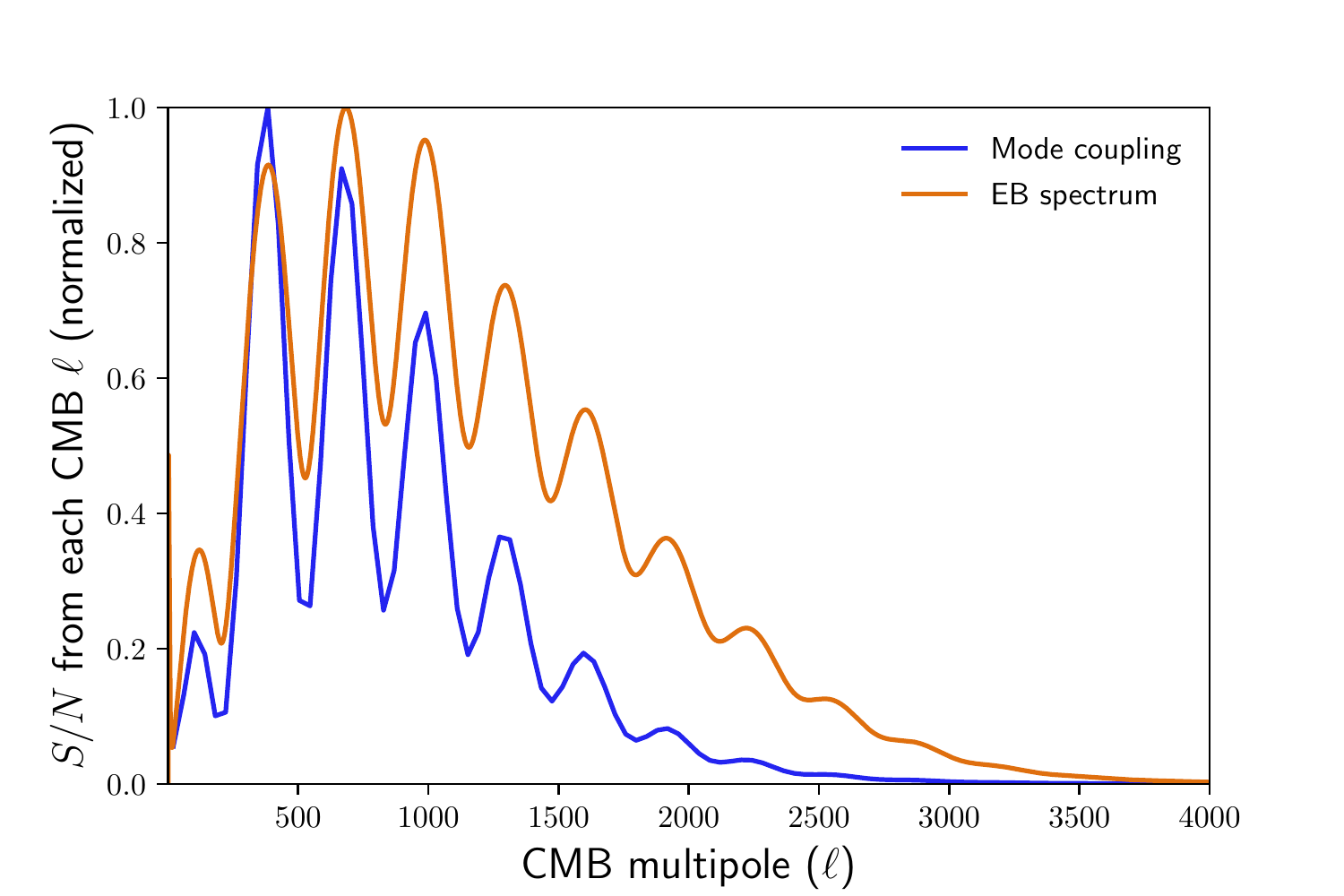}
\caption{
Contributions to the total signal-to-noise ratio from each CMB multipole, normalized so that the maximum value is unity. The ``mode coupling'' denotes the case using the cross-power spectrum between $\Img$ reconstructed from the $BB$ estimator and a perfectly correlated mass tracer. In the mode coupling case, we compute the reconstruction noise using CMB multipoles within a narrow multipole range whose center multipole and width are $\l$ and $\Delta \l=40$, respectively. For comparison, we also show the case using the $EB$ spectrum method. We assume the $5\,\mu$K-arcmin white noise for polarization with temperature noise $\sqrt{2}$ lower and the $1$ arcmin FWHM Gaussian beam. 
}
\label{fig:snrl}
\end{figure}

Here, we estimate sensitivity of a measurement of $\Img$ to the isotropic polarization angle. We consider a situation where the reconstructed $\Img$ is cross-correlated with a mass tracer of the large-scale structure, $m$, such as the cosmic infrared background (CIB) and galaxies. Because the cross-power spectrum probes $\beta C_L^{\grad \rm m}$ and $C_L^{\grad \rm m}$ is also observable, we can estimate the isotropic polarization angle, $\beta$. 

We compute the expected $1\,\sigma$ uncertainties of $\beta$ from a measurement of the cross-correlation between $\Img$ and a mass tracer, $m$, as:
\al{
    \sigma^{-2}(\beta) 
    &= \sum_L \frac{(2L+1)f_{\rm sky}}{(C_L^{\Img\Img}+N_L^{\Img\Img})(C_L^{\rm mm}+N_L^{\rm mm})+C_L^{\Img\rm m}}\left(\PD{C_L^{\Img\rm m}}{\beta}\right)^2
    \,, 
}
where $f_{\rm sky}$ is the observed sky fraction and $N_L^{\Img\Img}$ is the reconstruction noise spectrum. The reconstruction noise is dominated by the $BB$ estimator at lower noise experiments and $EB$ estimator at higher noise experiments. Thus, we evaluate a minimum variance reconstruction noise as $1/N_L^{\Img\Img}=1/A_L^{\Img,(EB)}+1/A_L^{\Img,(BB)}$. In the small angle limit, $\beta\ll 1$, the angular power spectra, $C_L^{\Img\Img}$ and $C_L^{\Img\rm m}$, are negligible in the denominator, and the above equation becomes:
\al{
    \sigma^{-2}(\beta) 
    &\simeq \sum_L \frac{(2L+1)f_{\rm sky}(C_L^{\grad\rm m})^2}{N_L^{\Img\Img}(C_L^{\rm mm}+N_L^{\rm mm})}
    \notag \\
    &= \sum_L (2L+1)f_{\rm sky}\rho^2_L\frac{C_L^{\grad\grad}}{N_L^{\Img\Img}}
    \,. \label{Eq:sigma-beta:mode}
}
Here, $\rho_L\equiv C_L^{\grad \rm m}/\sqrt{C^{\grad\grad}_L(C^{\rm mm}_L+N^{\rm mm}_L)}$ is the correlation coefficient between the true CMB lensing potential and the observed mass tracer. 

For comparison, we also compute the expected constraints on $\beta$ with the $EB$ spectrum method. The $1\,\sigma$ error on $\beta$ with the $EB$ spectrum method is given by:
\al{
    \sigma^{-2}(\beta) 
    &= \sum_\l \frac{(2\l+1)f_{\rm sky}}{(\CEE_\l+N^{\rm P}_\l)(\CBB_\l+N^{\rm P}_\l)+\CEB_\l}\left(\PD{\CEB_\l}{\beta}\right)^2
    \notag \\ 
    &\simeq 4\sum_\l \frac{(2\l+1)f_{\rm sky}(\CEE_\l)^2}{(\CEE_\l+N^{\rm P}_\l)(\CBB_\l+N^{\rm P}_\l)}
    \,, \label{Eq:sigma-beta:EB}
}
where $N^{\rm P}_\l$ is the CMB polarization noise power spectrum. 

Fig.~\ref{fig:comp} shows the expected constraints on $\beta$ given by \eq{Eq:sigma-beta:mode,Eq:sigma-beta:EB}. The constraints are shown as a function of the white noise level in polarization. We adopt the correlation coefficient between the CIB and true CMB lensing potential provided by \citet{Sherwin:2015}. For comparison, we also show an idealistic case where the true CMB lensing map is used as a mass tracer. If the noise level in polarization becomes lower than $\sim 3\mu$K-arcmin, the expected constraint on $\beta$ using the cross-power spectrum between $\Img$ and $m$ becomes close to that from the $EB$ spectrum method. 

Fig.~\ref{fig:snrl} shows contributions to the total signal-to-noise ratio (SNR) from each CMB multipole, normalized by the maximum value. Note that the SNR is proportional to $1/\sigma(\beta)$. We show the case using the cross-power spectrum between $\Img$ reconstructed from the $BB$ estimator and a perfectly correlated mass tracer. To see which CMB multipole is the most important for the SNR, we compute the reconstruction noise using CMB multipoles at each narrow multipole range whose center multipole is $\l$ and width is $\Delta \l=40$. For comparison, we also show the case with the $EB$ spectrum method. In the $EB$ spectrum method, we compute the SNR per multipole. The noise level and beam is the same as that of Fig.~\ref{fig:norm}. Compared to the $EB$ spectrum method, the SNR of the cross-power spectrum, $C_L^{\Img \rm m}$, comes from relatively large angular scales. 

Note that, for a low noise experiment in which the $B$-modes are dominated by the lensing-induced $B$-modes, we can apply delensing to reduce the lensing noise (e.g. \citealt{Kesden:2002:delens,Seljak:2003pn}). To see the impact of delensing on $\sigma(\beta)$, let us assume if the lensing contribution is partially removed by delensing and the delensed CMB map contains a residual lensing potential, $f_{\rm res}\grad$, with $0\leq f_{\rm res}\leq1$. In \eq{Eq:sigma-beta:mode}, the numerator scales as $f^2_{\rm res}$. On the other hand, the reconstruction noise of the $BB$ estimator depends on forth power of observed $B$-modes, and the denominator roughly scales as $f^4_{\rm res}$ because the lensing $B$-modes, the dominant source of the observed $B$-modes, are proportional to $f_{\rm res}\grad$. Thus, the right-hand-side of \eq{Eq:sigma-beta:mode} roughly scales as $1/f^2_{\rm res}$. Similarly, in \eq{Eq:sigma-beta:EB}, the right-hand-side scales as $1/f^2_{\rm res}$. From these intuition, delensing would not significantly change the difference of $\sigma(\beta)$ between the mode coupling and $EB$ spectrum methods. Note that the factor, $f_{\rm res}$, in general, depends on $\l$, although the $\l$-dependence is weak up to $\l\sim 2000$ (see e.g., \citealt{Namikawa:2017:delens,Namikawa:2018:nldelens}). An accurate evaluation of $\sigma(\beta)$ including delensing requires a more detailed analysis such as characterising delensing bias in small angular scales and is left for our future work. 


\section{Summary and Discussion} \label{sec:discussion}

We have explored the mode coupling involved in observed CMB anisotropies when the $\T B$ and $EB$ spectra are non-zero. We first derived the estimator for $\Img$ to extract the mode coupling introduced by lensing. We then generalized it to other mode-coupling cases. 
As an example of the application of the new mode coupling measurement, we explored the constraint on the isotropic polarization angle by reconstructing $\Img$. We showed that the polarization angle can be constrained by cross-correlating the reconstructed $\Img$ and mass tracers of the large-scale structure. The expected constraints on the polarization angle would be comparable to that obtained from the $EB$ spectrum method in future high precision polarization measurements such as CMB-S4 \citep{CMBS4} and LiteBIRD \citep{LiteBIRD}. 

The method discussed in this paper can be used to cross-check the isotropic polarization angle measured from the $EB$ spectrum method because the cross-power spectrum between $\Img$ and a mass tracer have a different response to observational systematics compared to the $EB$ spectrum. For example, the cross-power spectrum does not have a bias from the instrumental noise since the CMB noise and mass tracers does not correlate. Similarly, pointing errors could introduce a bias in the $EB$ spectrum \citep{Shimon:2007au} but do not in the cross-power spectrum with a mass tracer. The temperature-to-polarization leakage, on the other hand, could lead to a bias in the cross-power spectrum. The temperature-to-polarization leakage leads to non-zero correlations in the observed odd-parity spectra, $\T B$ and $EB$ \citep{P16:rot}. The kernel functions given in \eq{Eq:kernel:TE:Imx,Eq:kernel:TB:Imx,Eq:kernel:EE:Imx,Eq:kernel:EB:Imx,Eq:kernel:BB:Imx} then become non-zero even without the polarization rotation and introduce new mode-coupling terms in \eq{Eq:EE:lens-beta,Eq:EB:lens-beta,Eq:BB:lens-beta,Eq:TE:lens-beta,Eq:TB:lens-beta} which are proportional to $\grad$ as analogous to the isotropic polarization rotation. However, the kernel functions are different from that of the isotropic polarization rotation. Therefore, we can distinguish these contributions by constructing a bias-hardened estimator \citep{Namikawa:2012:bhe,Namikawa:2013:bhepol}. 

To constrain the cosmological signals of the isotropic cosmic birefringence, we need to mitigate an error of the polarization angle. The current accuracy of the polarization angle calibration is typically worse than $\sim 0.1$\,deg which is larger than the statistical errors in ongoing and future CMB experiments. Thus, we need to develop methods to avoid the large instrumental systematic bias in estimating the isotropic cosmic birefringence signals. In the $EB$ spectrum method, \citet{Minami:2019:rot} proposes to use the Galactic foregrounds which partially breaks the degeneracy between a bias from a miscalibration angle and signals of the isotropic cosmic birefringence. The method is already demonstrated with Planck data \citep{Minami:2020:planck}. Analogous to the $EB$ spectrum method, we can use the Galactic foregrounds involved in $\Img$. For example, the cross-power spectrum between $\Img$ and foreground-dominant frequency maps contains a Galactic foreground bispectrum. Although we need a model for the intrinsic Galactic foreground bispectrum, the joint measurements of the Galactic foreground bispectrum and $C_L^{\Img m}$ would break the degeneracy between a bias from an instrumental polarization angle error and the isotropic cosmic birefringence signal.

\section*{Acknowledgements}

TN thanks Blake Sherwin for helpful discussions. 
Some of the results in this paper have been derived using public software of the healpy \citep{healpy}, HEALPix \citep{Gorski:2004by}, and CAMB \citep{Lewis:1999bs}. 
For numerical calculations, this paper used resources of the National Energy Research Scientific Computing Center (NERSC), a U.S. Department of Energy Office of Science User Facility operated under Contract No. DE-AC02-05CH11231.
This work is also supported in part by JSPS KAKENHI Grant Number JP20H05859.

 



\bibliographystyle{mnras}
\bibliography{cite} 




\appendix




\bsp	
\label{lastpage}
\end{document}